\documentclass[a4paper,12pt,reqno,superscriptaddress,showkeys,nofootinbib]{revtex4}
\usepackage[centertags]{amsmath}
\usepackage{amsfonts}
\usepackage{amssymb}
\usepackage{amsthm}
\usepackage{newlfont}
\usepackage{stmaryrd}
\usepackage{mathrsfs}
\usepackage{mathtools}
\usepackage{euscript}
\usepackage{graphicx}
\usepackage{enumerate}
\usepackage{todonotes}
\usepackage{color}

\usepackage{graphicx}

\usepackage{amscd}
\usepackage{hyperref}

\theoremstyle{plain}

\theoremstyle{definition}

\theoremstyle{remark}

 \let\be=\beta

\newcommand{\opunit}{\text{1}\kern-0.22em\text{l}}

\DeclareMathAlphabet{\mathpzc}{OT1}{pzc}{m}{it}

\newcommand{\id}{\textrm{d}}

\begin{document}

\title{Frenetic bounds on the entropy production}

\author{Christian Maes\\
{\it Instituut voor Theoretische Fysica, KU Leuven}}

\begin{abstract} We show that under local detailed balance the expected entropy production rate is always  bounded in terms of the dynamical activity.  The activity refers to the time-symmetric contribution in the action functional for path-space probabilities and relates to escape rates and unoriented traffic.  Under global detailed balance we get a lower bound on the decrease of free energy which is known from gradient flow analysis. For stationary driven systems we recover some of the recently studied ``uncertainty'' relations for the entropy production, appearing in studies about the effectiveness of mesoscopic machines and that refine the positivity of the entropy production rate by providing lower bounds in terms of a positive and even function of the current(s).  We extend these lower bounds for the entropy production rate to include underdamped diffusions.
 \end{abstract}
 \maketitle

\begin{figure}[t]
\includegraphics[width=10 cm]{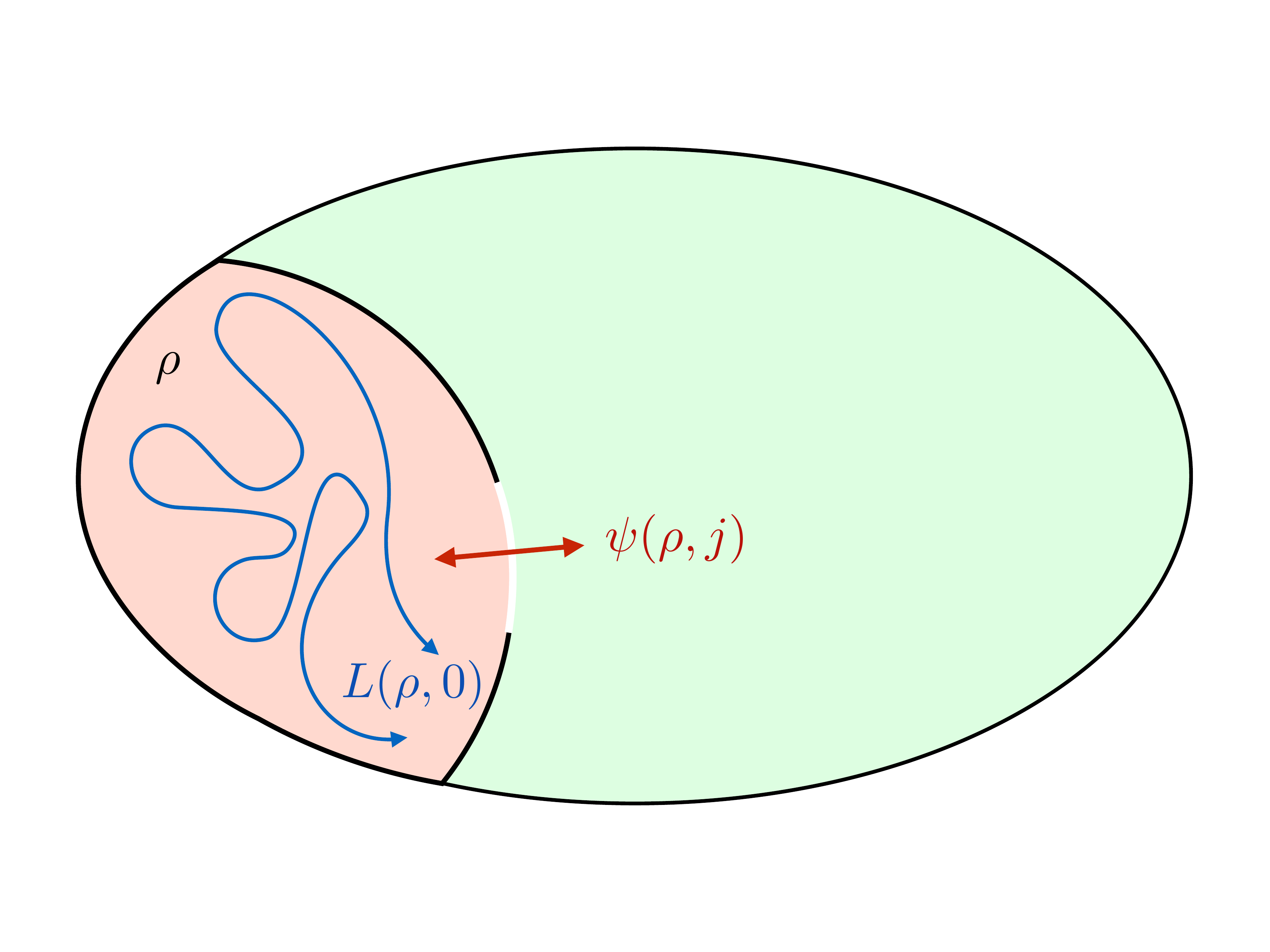}
\caption{\small{Entropy production is bounded from below by two types of activity.  For moving into a larger phase volume from condition $\rho$, thereby producing entropy, the dynamics must be sufficiently active inside $\rho$, represented by $L(\rho,0)$, and must readily find an escape route, represented by $\psi(\rho,j)$}.}\label{phspa}
\end{figure}

The expected entropy production rate $\sigma^*(\rho)$ given the present state $\rho$ of the system is always non-negative.  In this paper we provide lower bounds to $\sigma^*$ in terms of positive and even functions of the expected currents and forces.  These refinements of the second law give quantitative meaning to the saying that there can be no current without dissipation \cite{ncwh}, at least within the framework of dynamical fluctuation theory.  The latter was pioneered in the work of Onsager and Onsager-Machlup \cite{O,OM}.  More recently, Mielke, Peletier and Renger \cite{MPR13,pons} understood that detailed balance when expressed on the level of dynamical macroscopic fluctuations,  implies a (possibly nonlinear) gradient flow in the relaxation to equilibrium.  Here we use similar ideas to obtain bounds on $\sigma^*$ not only for the return to equilibrium in which case it implies bounds on equilibration times, but also for driven systems, either in the steady state or in the relaxation to the nonequilibrium steady condition. In that way the present results connect with \cite{udo,udo3,udo4,mit,udo2,mit2,laz,hal,cha} where inequalities for currents and entropy production have appeared in stationary nonequilibria, mostly in the context of mesoscopic machine efficiency. Our arguments provide unification, and physically point to what really matters in all that, which is the dynamical activity allowing the production of entropy.\\
  Heuristically, dynamical activity  indicates the time-symmetric changes, both in the exit/access of states as in being inversely proportional to the residence or sojourn time in a macroscopic condition.  We have called it the {\it traffic} before, \cite{traf}, and in response theory we have been speaking about the {\it frenetic} contribution \cite{fdr,njp}, for example for the understanding of negative differential conductivity \cite{kolk}.  In Boltzmann's picture of relaxation to equilibrium in terms of a flow on phase space visiting the various regions of macro-states, it adds ``surface'' to ``volume'' considerations, as it refers to the accessibility of and the escape-ability from macroscopic conditions, and not only to their entropy; see the cartoon in Fig.~\ref{phspa}.

The starting point of our analysis is dynamical fluctuation theory, \cite{com,revjona,DV}. That means we consider the probability of empirically possible trajectories, specified at times $s\in [0,t]$ via a variable density $\rho_s$ and a current $j_s$.   Those $\rho_s$ can be fields on physical space or they can be empirical averages of particle properties {\it etc}. The empirical rate of change in time is given by the current $j_s$ which often obeys a further constraint such as in the continuity equation $\dot{\rho_s} + {\cal D} j_{\rho_s} = 0$ where $\cal D$ is a divergence, but other operators $\cal D$ are possible in case $\rho_s$ does not refer to a locally conserved quantity. We repeat that $(\rho_s,j_s), s\in [0,t],$ denotes a \emph{possible} density-current trajectory on some level of coarse graining, realizable from the microscopic laws, constituents and initial conditions.
The probability of such a trajectory is then written as
\begin{equation}\label{prob}
\text{Prob}[(\rho_s,j_s), 0\leq s\leq t] \simeq e^{N S(\rho_0)}\,e^{-N\int_0^t\id s\, L(\rho_s,j_s)}
\end{equation}
Here $N \uparrow +\infty$ is a scale parameter like the number of independent copies of a finite Markov process, or the number of particles, the volume, {\it etc.} where \eqref{prob} must be understood in the sense of fluctuations around a law of large numbers, {\it i.e.}, giving the asymptotic exponential behavior of probabilities.  In fact, \eqref{prob} is the formal extension of Boltzmann--Planck--Einstein macroscopic fluctuation theory to the time-domain \cite{com,revjona}, where now the trajectory $(\rho_s,j_s), s\in [0,t],$ is the random variable. The functional $S$ would be a thermodynamic potential (entropy per $k_B$) when the Prob in \eqref{prob} refers to the thermodynamic equilibrium distribution; if in stationary nonequilibrium $-S$ is still sometimes called a nonequilibrium free energy to emphasize the analogy.   The Lagrangian $L$ in \eqref{prob} further determines the plausibility of the various possible trajectories.   Note that apart from the structural aspects, the Lagrangian $L$ in \eqref{prob} is essentially constructed and explicitly known when starting from a (semi-)Markov process, \cite{lag1,lag2,jona,epl,semi}.  That is part of the mathematical work that started with the large deviation theory for Markov processes in \cite{DV,FW,fen}. Properties that we assume from the outset are that $L\geq 0$ and is convex in $j$ for all $\rho$. Note that due to the large $N$ the weights in \eqref{prob} are exponentially small with respect to the zero--cost flow $j_s^* = j^*(\rho_s)$ which solves $L(\rho_s,j_s^*)=0$ for all times $s$  and induces a unique evolution equation $\dot{\rho_s} + {\cal D} j^*_s=0$.  Under the present assumptions, finding the typical trajectory is thus equivalent with finding for any  given $\rho$ the solution $j^*=j^*(\rho)$ of $L(\rho,j^*)=0$. To that $j^*$ corresponds the expected entropy production rate $\sigma^*=\sigma(\rho,j^*)$ for which we want to obtain bounds as the main goal of this paper.  Remark that the unit of time is arbitrary here, and currents (with dimension of frequency) must always be compared with other time-constants.

Even though we often refer to macroscopic conditions it is important to remember that the theory also applies to small systems, which are either mentally copied $N$ times or for which the motion is repeated $N$ times. For example, the Master equation for mesoscopic dynamics as {\it e.g.} finite state continuous time Markov jump processes, is the macroscopic evolution equation governing the density of independent random walkers on the graph of states with corresponding transition rates. {\it E.g.} for Markov jump processes the zero-cost flow $j^*(\rho_s)$ is the expected current appearing in the Master equation. Yet the present theory is not restricted to Markov processes, nor to linear macroscopic equations.  Semi-Markov processes also allow a physically interesting dynamical fluctuation theory \cite{semi}, and path-space large deviations are also well-known for certain nonlinear evolutions \cite{bou}.  Another novelty of the present approach is that we include underdamped dynamics where some variables (velocities) are odd under kinematic time-reversal.\\

{\it Entropy production and dynamical activity.}--- 
An important question in the construction of nonequilibrium statistical mechanics is to identify the phenomenological and operational meaning of  the Lagrangian $L(\rho,j)$ in \eqref{prob}.   For that purpose $L$ is most usefully split up in a time-symmetric and a time-antisymmetric part,
\begin{equation}\label{ep}
L(\rho,j) = \frac 1{2}[L(\rho,j) + L(\rho,-j)] - \frac 1{2}\sigma(\rho,j)
\end{equation}
where $\sigma(\rho,j) = L(\rho,-j) - L(\rho,j)$ is antisymmetric under time-reversal.  When the density $\rho$ is determined by time-symmetric variables (being even under kinematic time-reversal), $\sigma(\rho,j)$ is the entropy production rate per $k_B$ corresponding to the couple $(\rho,j)$.  Entropy production rate is a bilinear form of forces and fluxes/currents.  That identification of \eqref{ep} with entropy production rate follows from the condition of local detailed balance as holds for externally driven systems, or for systems that are in weak contact with multiple equilibrium reservoirs that are spatio-temporally sufficiently separated, {\it cf.} \cite{ldb2,ldb3}.\\
  In the context of traditional irreversible thermodynamics, without external driving and under local equilibrium,
\begin{eqnarray}\label{fe}
\sigma(\rho,j) &=& \int_\Omega\id X \,\nabla\frac{\delta S}{\delta \rho(X)}\cdot j(X) = \int_\Omega \id X\,\frac{\delta S}{\delta \rho(X)}\dot{\rho}(X) \nonumber\\
&= & \frac{\id}{\id t}S[\rho(X), X\in \Omega]
\end{eqnarray}
where we took $\rho(X),j(X)$ on some spatial domain $X\in \Omega$ with periodic boundary conditions, and $\nabla j = - \dot{\rho}$ for a locally conserved field. We then have that for all possible $(\rho,j)$ the (variable) entropy production rate is the time-derivative of the entropy $S$ and the thermodynamic force is the gradient of its variational derivative, but of course the expected entropy production rate given $\rho$ requires knowing the expected current $j^*(\rho)$.\\ 
In the case of steady driven systems, 
\begin{equation}\label{dc}
\sigma(\rho,j) = \int\id X \,F(\rho(X),X)\cdot j(X)
\end{equation}
where the force $F$ is now not globally derivable from a thermodynamic potential.   (In our units where entropy is measured per $k_B$ we take the energy dimensionless.)  At any rate, the entropy production rate $\sigma(\rho,j)$ is linear in the current $j$ for the given force $F$ that depends on the local density $\rho$. 

We continue with the time-symmetric part of the Lagrangian,
\begin{equation}\label{tsc}
L(\rho,j) + L(\rho,-j) = 2\psi(\rho,j) + 2L(\rho,0)
\end{equation}
There are two contributions, both non-negative as will turn out.  The $L(\rho,0)\geq 0$ is an (internal) dynamical activity, when there is no change in the macroscopic condition ($j=0$).  It corresponds to the activity of the more microscopic degrees of freedom expected at $j=0$. $L(\rho,0)$ is the escape rate from condition $\rho$.  On the other hand the $\psi(\rho,j)$ gives the dynamical activity corresponding to a given current $j\neq 0$; see Fig.~\ref{phspa}. We have of course $\psi(\rho,j)=\psi(\rho,-j)$, symmetric in $j$ with $\psi(\rho,0)=0$, and thus corresponds to the unoriented traffic between ``neighboring'' conditions.  Moreover, as by construction,
\begin{equation}\label{ps}
\psi(\rho,j) = L(\rho,j) - L(\rho,0) + \frac 1{2}\sigma(\rho,j)
\end{equation}
it follows that $\psi(\rho,j)$ is convex in $j$, which implies that $\psi(\rho,j)\geq 0$.\\   By the time-symmetry (always assuming even variables), both $\psi(\rho,j)$ and $L(\rho,0)$ stand for a traffic-like quantity which is invariant under time-reversal \cite{traf,kolk}.  Together, depicted in Fig.~\ref{phspa}, they inform us about the activity in the system at $(\rho,j)$.\\
It is easy to check that writing $\sigma(\rho,j) = F\cdot j = L(\rho,-j) - L(\rho,j)$,
\[
2L(\rho,0) = \sup_j\,\{F\cdot j - 2\psi(\rho,j)\}
\]
from substituting \eqref{ps}.  Therefore, $0\leq L(\rho,0) = \hat\psi(\rho,F)$ is the Legendre transform of $\psi(\rho,\cdot)$ at force $F$ and the decomposition \eqref{tsc} is in a pair of convex duals, with current and force being conjugate.  For a dynamics with force $F$, the Lagrangian \eqref{ep} has therefore the canonical structure
\begin{equation}\label{eps}
L(\rho,j) = \psi(\rho,j) + \hat\psi(\rho,F) - \frac 1{2} \sigma(\rho,j)
\end{equation}
Such a structure of joint density-current fluctuations away from equilibrium was developed in \cite{lag1,lag2,epl} to which we refer for more discussion.\\

{\it Zero-cost flow}---
When the current $j=j^*$ minimizes the action, $L(\rho,j^*)=0$ (zero cost), \eqref{eps} returns the expected entropy production rate $\sigma^* = \sigma(\rho,j^*(\rho)) = j^*(\rho)\cdot F$, function of condition $\rho$ and force $F$, as
\begin{equation}
\label{0cf}
\frac 1{2} \,\sigma^* = \psi(\rho,j^*) + \hat\psi(\rho,F)
\end{equation}
In words, when we are given macroscopic condition $\rho$, then the expected and in fact most likely fate of the system is to take the current $j^*(\rho)$ for which the entropy production rate equals the dynamical activity as in \eqref{0cf}. Macroscopic trajectories can be characterized as those for which at each moment there is a perfect balance between the entropy production rate and the dynamical activity given the present state.  As an immediate consequence follows that
 every lower bound on the dynamical activity (right-hand side of \eqref{0cf}) implies a lower bound on the expected entropy production rate, which is the first main result, a frenetic bound on the entropy production rate.  In particular,
it realizes quantitatively the statement that there cannot be
a current without entropy production. The right-hand side is indeed strictly positive from
the moment $j^*\neq 0$. It also says that the expected entropy production rate is universally
bounded from below by the dynamical activity related to the expected current.  In that respect it is useful to know that the driving force $F$ in \eqref{0cf} can  be re-expressed in terms of the expected current $j^*$, through the equation $F(\rho,j^*) = \partial_j \psi(\rho,j^*/2)$ which follows from \eqref{eps}.

We can also derive a number of
further inequalities which naturally follow from positive lower bounds $\psi(\rho,j)\geq 0$ and $\psi^*(\rho,F)\geq 0$.  Both are convex functions for which the Taylor theorem with remainder gives a first simplest bound. Their Hessian matrix is strictly positive in all finite regions around current $j=0$ (for $\psi(\rho,j)$) or around forcing $G=0$ (for $\hat\psi(\rho,G)$).  It is
physically natural to assume that there is saturation in both currents and in forces in the sense that there is a maximal possible current amplitude and that all possible forces have a maximal strength.  We define then the
time-constant $m(\rho)$  and frequency $\kappa(\rho)$ from the Hessian matrix of $\psi$ with respect to currents $j$, respectively of $\psi^*$ with respect to forces $G$: in the sense of quadratic forms,
\[ 
m(\rho) = \inf_j \text{Hess}(\psi(\rho,j)),\quad 
\kappa(\rho) = \inf_G \text{Hess}(\hat\psi(\rho,G))
\] 
where the first infimum is over all currents $j$ with amplitude below $|j|\leq j_\text{max}$, a saturation or maximal possible current amplitude $j_\text{max}$. Similarly, the second infimum is over all possible forces $G$ which in amplitude are below a maximum $G_\text{max}$.  For our second main result,
\[ 
\psi(\rho,j^*) \geq \frac 1{2}m(\rho)\,(j^*)^2,\quad L(\rho,0) = \hat\psi(\rho,F) \geq \frac 1{2} \kappa(\rho)\,F^2
\] 
and hence
\begin{equation}\label{main2}
 \sigma^* \geq  m(\rho)\,(j^*)^2 + \kappa(\rho)\,F^2
\end{equation}
for $m(\rho),\kappa(\rho) >0$ which measure the quadratic growth of the dynamical activity near $j=0$, respectively $G=0$.
Remark that \eqref{main2}  is the simplest general bound, naturally following from convexity and allowing explicit calculations as we will show. Yet, compared to
the universal \eqref{0cf}, the bounds \eqref{main2} are in the spirit of close-to-equilibrium where
the entropy production rate indeed becomes quadratic in currents or in forces. For every  application there may well be better or more natural bounds and each time \eqref{0cf} carries the promise that those yield bounds on the entropy production rate. 
On the other hand, any bound on the expected entropy production rate $\sigma^*$ implies a bound on the dynamical activity.\\

{\it Examples}---
There are various classes of examples and we will not treat the more standard ones.  {\it E.g.}, it is well-documented that the dissipative aspects of return to thermodynamic equilibrium as described via macroscopic equations satisfy gradient flow \cite{pons}.  We imagine the evolution thus as follows.  Given $\rho$ the system follows steepest descent on the free energy landscape to find the ``next'' condition.  What is steepest obviously depends on the local curvature and that is determined by a metric in the space of those macroscopic conditions.  The metric is provided by the positive definite Onsager matrix for the return to equilibrium.  In fact, the Ricci curvature in the space of densities for macroscopic equations like the nonlinear diffusion equation gives a bound on the relaxation time \cite{vil}. We now understand these results physically via lower bounds on the entropy production rate. For close-to-equilibrium processes or for driven diffusive systems holds the quadratic form
\begin{equation}\label{psir}
\psi(\rho,j) = \frac 1{4}\int \id X \,j(X)\cdot \Gamma^{-1}(\rho(X))\,j(X)
\end{equation}
for a symmetric Onsager-matrix $\Gamma$ which depends on the local field.  From \eqref{psir} it is clear that bounds on the Onsager matrix provide bounds on the dynamical activity, hence, from \eqref{0cf} bounds on the free energy decay \eqref{fe}.\\
The same remains true for overdamped Markov diffusion processes, driven or not.  Consider 
a  $d-$dimensional inhomogeneous diffusion, with It\^o convention,
\[ 
  \dot{x}_t = \chi(x_t) \, f(x_t)  + \nabla\cdot D(x_t)
  + \sqrt{2 D(x_t)}\, \xi_t
\] 
The mobility $\chi(x) = \be D(x)$ is a positive $d\times d-$matrix, $\be > 0$ is the inverse bath-temperature and $\xi_t$ is standard white noise.
Under smoothness and confining boundary conditions  the Lagrangian is obtained in \cite{lag2},
and 
for $\psi(\rho,j)$ we get exactly
\eqref{psir} with $\Gamma =\rho D$ as indeed the quadratic form is already exact here.  Furthermore, $L(\rho,0) = \psi(\rho,j^*)$ so that from \eqref{0cf} the expected entropy production rate trivially obtains the lower bound,
\begin{equation}\label{ddsb}
\sigma^* \geq \frac 1{\max_x{||D(x)||}} \int (j^*(x))^2\id x
\end{equation}

Things become more interesting for jump processes where nonlinearities are more prominent. Yet all we need remains explicit and we can use the formul{\ae} 4.1--4.7 in \cite{lag1} to find that for a general Markov jump process with transition rates $\lambda(x,y)$ between states $x\rightarrow y$,
\begin{eqnarray}
L(\rho,0) &=& \frac 1{2}\sum_{x\neq y}\frac{(j^*(x,y))^2}{(\sqrt{\rho(x)\lambda(x,y)}+\sqrt{\rho(y)\lambda(y,x)})^2}
\nonumber\\
\psi(\rho,j) &=& \frac 1{4}\sum_{x\neq y}[ j(x,y)\log \frac{j(x,y) + \sqrt{j^2(x,y) + \gamma(x,y)}}{-j(x,y) + \sqrt{j^2(x,y) + \gamma(x,y)}}\nonumber \\ &&+ 2\sqrt{\gamma(x,y)} - 2\sqrt{j^2(x,y) + \gamma(x,y)}]
\end{eqnarray}
for $\gamma(x,y) = 4\rho(x)\rho(y)\lambda(x,y)\lambda(y,x) = \gamma(y,x)$ measuring time-symmetric reactivity.  It is easy enough to check that
\begin{eqnarray}
\psi(\rho,j) 
&\geq& \frac{\alpha}{2}\,\sum_{x\neq y}\frac{j^2(x,y)}{\sqrt{j^2(x,y) + \gamma(x,y)}}\\
&\geq& \frac{\alpha^2}{2(1-\alpha)}\,\sum_{x\neq y}\frac{j^2(x,y)}{\sqrt{\gamma(x,y)}}\nonumber
\end{eqnarray}
whenever $j^2(x,y) \leq (1-2\alpha)\,\gamma(x,y)/\alpha^2$.  
Thence, \eqref{0cf} gives a lower bound for the expected entropy production rate, as
\begin{eqnarray}\label{sb}
\sigma^* &\geq&  \frac{\alpha^2}{2(1-\alpha)}\,\sum_{x\neq y}\frac{(j^*(x,y))^2}{\sqrt{\rho(x)\rho(y)\lambda(x,y)\lambda(y,x)}}\nonumber\\
&+& \sum_{x\neq y}\frac{(j^*(x,y))^2}{(\sqrt{\rho(x)\lambda(x,y)}+\sqrt{\rho(y)\lambda(y,x)})^2}
\end{eqnarray}
when the current between $x\rightarrow y$ can never exceed the saturation amplitude $j_\text{max}(x,y) =  (1-2\alpha)\,\gamma(x,y)/\alpha^2$ written in terms of a parameter $\alpha<1/2$.  Of course, \eqref{sb} holds also for each $(x,y)-$term separately as lower bound.\\
As example take a quantum dot with states
$x = 0,1$, which we connect  to a left and a right load in the Coulomb blockade regime.  In chemical language, there are two channels for each transition $x \rightarrow 1-x$. Put $F\geq 0$ (constant) for the potential gradient so that local detailed balance at inverse temperature $\beta$ demands transition rates corresponding to each channel of the
following general form,
\begin{equation}\label{eq: example}
\begin{array}{ll}
  k_L(0,1) = \nu\, e^{\be F / 2}, & k_L(1,0) = \nu\, e^{-\be F / 2} \\
  k_R(0,1) = \nu\, e^{-\be F / 2}, & k_R(1,0) = \nu\, e^{\be F / 2}
\end{array}
\end{equation}
where $\nu=\nu(\beta,F)$ is a reference frequency.
We write the densities as
$\rho(0) = p = 1-\rho(1)$ and the current $j=j(0,1)$ goes from left to right. The Lagrangian was obtained in \cite{epl}.
A computation from \eqref{tsc} yields
\begin{eqnarray}
L(p,0) &= & 2 \nu \Bigl[ \cosh\frac{\be F}{2} -
  \sqrt{4p(1-p) } \,\Bigr] \nonumber\\
  \psi(p,j) &=& j\log\frac{j + \sqrt{4p(1-p)\nu^2 + j^2}}{-j+ \sqrt{4p(1-p)\nu^2 + j^2}}\label{qlb}\nonumber\\ &+& 2\left[\sqrt{4p(1-p)\nu^2} - \sqrt{4p(1-p)\nu^2 + j^2}\right]\nonumber
  \end{eqnarray}
We have $L(p,0) 
  \geq \frac 1{2}\,\nu\,(\beta F)^2$, and  $
  \frac{\partial^2}{\partial j^2}\,\psi(p,j) = \frac{2}{\nu}\,\frac 1{\sqrt{4p(1-p)+(j/\nu)^2}}$
leads to the time-constant $m(p) = 2[4p(1-p)+\alpha^2]^{-1/2}/\nu$ in the range $|j/\nu|\leq \alpha$. Therefore, for \eqref{main2}, 
  \begin{equation}\label{s2}
  \sigma^* \geq \frac{2}{\nu}\,\frac 1{\sqrt{4p(1-p)+\nu^2j_\text{max}^2}} (j^*)^2 + \beta^2\nu\, F^2 
  \end{equation}

  A case which is absent in the literature for bounds on the entropy production is that of an underdamped Markov diffusion process.  
For processes with degrees of freedom which are odd under kinematic time-reversal there exists obviously a condition of generalized detailed balance but the arguments above cannot be applied as such.  Instead we need the relation \eqref{ep} but for fluctuations around the Hamiltonian flow.\\
Suppose indeed quite more general than \eqref{ep} that there is a function $h_\rho$ so that the Lagrangian satisfies
\begin{equation}\label{tfo}
L(\rho,h_\rho-j) - L(\rho,h_\rho + j) =  \sigma(\rho,j)
\end{equation}
where $\sigma(\rho,j)$ is as before the physical entropy production rate corresponding to the pair $(\rho,j)$, \cite{gen1}. Then, the analysis of the previous sections can be repeated exactly by replacing there $L$ with $\tilde{L}(\rho,j) = L(\rho, j+ h_\rho)$.  The new Lagrangian $\tilde L$ inherits all the properties of $L$, and gives rise to dynamical activity $\tilde{\psi}(\rho,j), \tilde{L}(\rho,0)$ as before.  We thus obtain bounds on the entropy production rate $\sigma^*$ from the same reasonings as above.  The simplest quadratic scenario is
\[
L(\rho,j) = \frac 1{4} (j-h_\rho- AF(\rho))\cdot\frac 1{\rho\,A} (j-h_\rho- AF(\rho))
\]
for force $F$ and matrix $A$.  Then indeed, $\tilde{L}(\rho,-j)- \tilde{L}(\rho,j) = j\cdot F(\rho)$ is the physical entropy production rate.
For example, for the Kramers equation with energy
$E(q,p) = p^2/2 + U(q)$ and with symplectic matrix $K$, we have $h_\rho = -\rho K[\nabla E + (-f(q),0)^T]$ and $F(\rho) = -\rho \,[\nabla(E + \log \rho)+ (-f(q),0)^T]$, with $\nabla = (\partial_q,\partial_p)$. Then,  the symmetry \eqref{tfo} holds for the dynamics $\dot{q} =p, \dot{p} = -\partial_q U(q) + f(q) - \gamma\,p + \sqrt{2\gamma}\,\xi_t$,
with $\xi_t$  standard white noise and for $A$ having matrix element $A_{pp}=\gamma$ and zeroes elsewhere.\\

{\it Conclusion}---
Lower bounds on the expected entropy production rate are obtained in terms of time-symmetric activity.  The theory of dynamical macroscopic fluctuations provides a unifying framework for these refinements of the second law, and extends previous work on thermodynamic uncertainties to underdamped and possibly nonMarkov processes.

\end{document}